\documentclass[aps,natphysics,superscriptaddress,twocolumn,amsmath,amssymb, balance,notitlepage,bibnotes]{revtex4-2}
\usepackage{xcolor,bm,multirow,times}
\usepackage{physics}
\usepackage{rotating}
\usepackage{verbatim}
\usepackage{graphicx,tabularx}
\usepackage[sort&compress]{natbib}
\usepackage[T1]{fontenc}
\usepackage{soul}
\usepackage{dcolumn}
\usepackage{bm}
\usepackage[version=4]{mhchem}
\usepackage{booktabs}
\usepackage{gensymb}
\usepackage{color}
\usepackage[colorlinks=true,urlcolor=blue,linkcolor=magenta,citecolor=magenta]{hyperref}
\usepackage{enumitem}

\newcommand{\kcso}{K$_2$Co(SeO$_3$)$_2$}

\allowdisplaybreaks

\begin{document}
\title{Sharp spectroscopic fingerprints of disorder in an incompressible magnetic state}
\author{Chaebin Kim}
\thanks{These authors contributed equally to this work.}
\affiliation{School of Physics, Georgia Institute of Technology, Atlanta, Georgia 30332, USA}
\author{Sumedh Rathi}
\thanks{These authors contributed equally to this work.}
\affiliation{School of Physics, Georgia Institute of Technology, Atlanta, Georgia 30332, USA}
\author{Naipeng Zhang}
\affiliation{National High Magnetic Field Laboratory, Tallahassee, Florida 32310, USA}
\author{Arnab Seth}
\affiliation{School of Physics, Georgia Institute of Technology, Atlanta, Georgia 30332, USA}
\author{Nikolai V. Simonov}
\affiliation{School of Physics, Georgia Institute of Technology, Atlanta, Georgia 30332, USA}
\author{Aya Rutherford}
\affiliation{Department of Physics and Astronomy, University of Tennessee, Knoxville, Tennessee 37996, USA}
\author{Long Chen}
\affiliation{Department of Physics and Astronomy, University of Tennessee, Knoxville, Tennessee 37996, USA}
\author{Haidong Zhou}
\affiliation{Department of Physics and Astronomy, University of Tennessee, Knoxville, Tennessee 37996, USA}
\author{Cheng Peng}
\affiliation{Department of Chemistry, Michigan State University, East Lansing, Michigan 48864, USA}
\author{Mingyu Xu}
\affiliation{Department of Chemistry, Michigan State University, East Lansing, Michigan 48864, USA}
\author{Weiwei Xie}
\affiliation{Department of Chemistry, Michigan State University, East Lansing, Michigan 48864, USA}
\author{Advik D. Vira}
\affiliation{School of Physics, Georgia Institute of Technology, Atlanta, Georgia 30332, USA}
\author{Mengkun Tian}
\affiliation{Institute for Matter and Systems, Georgia Institute of Technology, Atlanta, Georgia 30332, USA}
\author{Mykhaylo Ozerov}
\affiliation{National High Magnetic Field Laboratory, Tallahassee, Florida 32310, USA}
\author{Itamar Kimchi}
\email{ikimchi3@gatech.edu}
\affiliation{School of Physics, Georgia Institute of Technology, Atlanta, Georgia 30332, USA}
\author{Martin Mourigal}
\email{mourigal@gatech.edu}
\affiliation{School of Physics, Georgia Institute of Technology, Atlanta, Georgia 30332, USA}
\author{Dmitry Smirnov}
\email{smirnov@magnet.fsu.edu}
\affiliation{National High Magnetic Field Laboratory, Tallahassee, Florida 32310, USA}
\author{Zhigang Jiang}
\email{zhigang.jiang@physics.gatech.edu\\}
\affiliation{School of Physics, Georgia Institute of Technology, Atlanta, Georgia 30332, USA}

\date{\today}

\maketitle

{\bf Disorder significantly impacts the electronic properties of conducting quantum materials by inducing electron localization and thus altering the local density of states and electric transport. In insulating quantum magnetic materials, the effects of disorder are less understood and can drastically impact fluctuating spin states like quantum spin liquids. In the absence of transport tools, disorder is typically characterized using chemical methods or by semi-classical modeling of spin dynamics. This requires high magnetic fields that may not always be accessible. Here, we show that magnetization plateaus---incompressible states found in many quantum magnets---provide an exquisite platform to uncover small amounts of disorder, regardless of the origin of the plateau. Using optical magneto-spectroscopy on the Ising-Heisenberg triangular-lattice antiferromagnet K$_2$Co(SeO$_3$)$_2$ exhibiting a 1/3 magnetization plateau, we identify sharp spectroscopic lines, the fine structure of which serves as a hallmark signature of disorder. Through analytical and numerical modeling, we show that these fingerprints not only enable us to quantify minute amounts of disorder but also reveal its nature---as dilute vacancies. Remarkably, this model explains all details of the thermomagnetic response of our system, including the existence of multiple plateaus. Our findings provide a new approach to identifying disorder in quantum magnets.}

\section*{Introduction}
Chemical disorder is a fundamental characteristic of solid-state materials and can manifest in various ways, such as impurities and dopants, site mixing and vacancies, bond disorder, and topological defects. In quantum materials, this disorder is imprinted at high temperatures during synthesis and becomes kinetically frozen, often remaining unnoticed until it influences putative emergent phenomena at low temperatures. However, disorder is not always detrimental in quantum systems. A notable example is the quantum Hall effect (QHE), where disorder plays a crucial role in the formation of quantized plateaus in the Hall resistance of devices and materials. This phenomenon has led to the establishment of a new international standard for electrical resistance~\cite{Jeckelmann_2001,Klitzing_2005}. 

The role of disorder in quantum \textit{magnetic} materials is less understood yet it is crucial for a wide range of systems, particularly quantum spin liquids (QSLs). These systems are believed to host topological order and itinerant fractional excitations without exhibiting clear, conventional order-parameter-like signatures. The response, survival, and nature of potential QSLs in the presence of chemical disorder are not fully understood. Increasingly, researchers are recognizing that disorder can mimic the expected behavior of a QSL~\cite{Zhu2017, KimchiSenthil2018, KimchiLee2018,Kao2021,syzranov2022, Lane2025}. But a significant challenge lies in quantifying the level of disorder using chemical techniques and translating this information into a model magnetic Hamiltonian, $\mathcal{H}$, whose emergent quantum many-body phases are not known in advance. A successful approach involves employing electron spin resonance \cite{komatsu1996, miksch2021} or fitting inelastic neutron scattering and optical spectroscopy measurements under sufficiently high magnetic fields to polarize the system. In this semi-classical regime, it becomes possible to directly measure  $\mathcal{H}$~\cite{Coldea2002}, including the level of chemical disorder~\cite{Li2017,Steinhardt2021}. However, practical limitations arise from the need for large magnetic fields. Furthermore, spin-anisotropic systems of current interest may only reach saturation asymptotically~\cite {Zhou2023}. 

The complex magnetization process of antiferromagnets may offer an alternative strategy. By utilizing quantized magnetization plateaus---when they occur---we show that it is not only possible to extract $\mathcal{H}$ at a much lower magnetic field, but these plateaus also exhibit unique, sharp spectroscopic signatures allowing for the detection of disorder with high precision. Similar to the QHE, this sensitivity emerges from the distinctive structure of incompressible magnetic states with quantized magnetization, an effect which has not been widely acknowledged until now.

\begin{figure*}[htb!]
        \centering
        \includegraphics[width=1.0\linewidth]{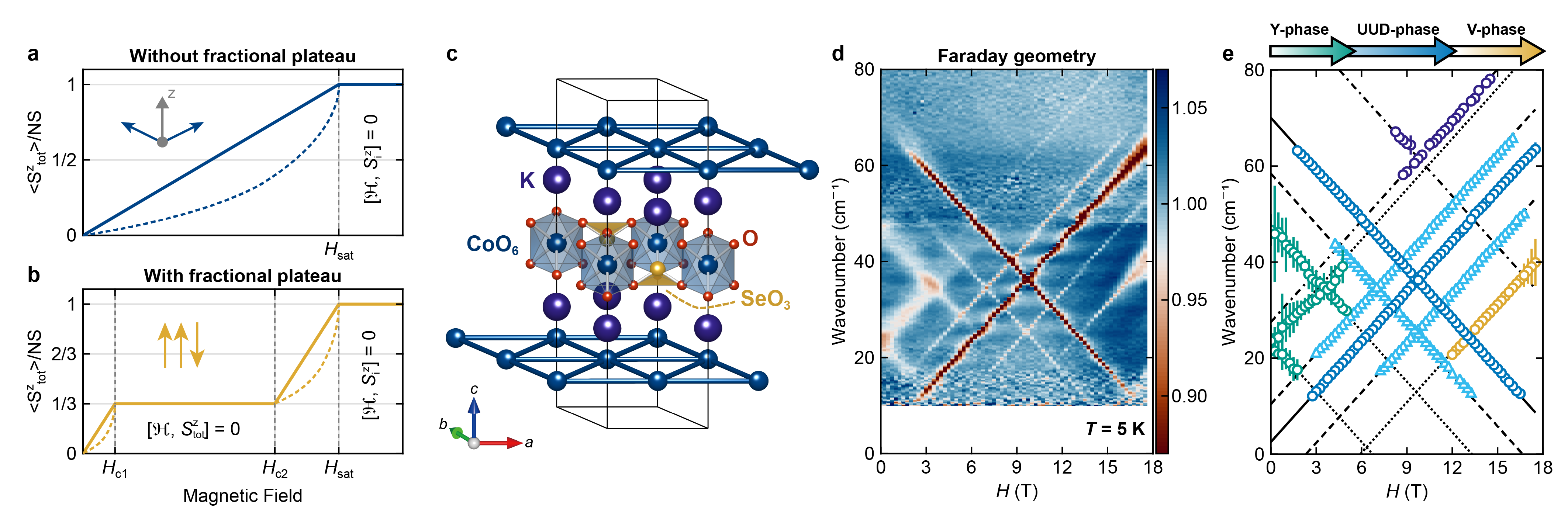}
        \caption{\textbf{Magnetization plateau and far-infrared magneto-spectroscopy measurements of \kcso.} \textbf{a, } Schematic representation of the normalized uniform magnetization $m=\langle S_{\rm tot}^z \rangle/NS$ of an ordered antiferromagnet described at the classical level (solid line) or including quantum corrections (dashed line). The spin-$S$ moments form a canted structure and become fully polarized at the saturation magnetic field $H_{\rm sat}$. \textbf{b, } Normalized uniform magnetization of an antiferromagnet with a field-induced quantized plateau state, which is stabilized between $H_{\rm c1}$ and $H_{\rm c2}$. The UUD state is a possible classical state realizing a $m=1/3$ magnetization plateau. \textbf{c, } Crystal structure of KCSO. Orange spheres represent oxygen, purple represents potassium, blue represents cobalt, and yellow represents selenium ions. The triangular lattice of magnetic Co$^{2+}$ ions is shown by blue bonds. \textbf{d, } Frequency and magnetic field dependence of the normalized magneto-transmission spectra $I(\omega; H)$ of KCSO measured in Faraday geometry at $T\!=\!5$~K with the magnetic field direction $z$ along the $c$-axis, see Methods and Supplementary Sec. IA-B for details of our measurement and data analysis protocols. \textbf{e, } Energy modes $E(H)$ extracted from Lorentzian fits to the spectra at each magnetic field. Colors encode the nature of the underlying magnetic state: green and yellow for the compressible Y-like and V-like supersolid phases, and blue, light blue, and purple for the incompressible UUD phase corresponding to a $m=1/3$ plateau. Black lines are linear fits to the data using the averaged $g_{z}$ values, see Supplementary Sec. IC for details. Dotted lines represent excitations originating from the Y and V phases, solid lines indicate dominant spin-flip modes, dot-dashed lines show the triple spin-flip modes, and dashed lines correspond to disorder-induced satellite modes of the UUD phase.}
        \label{fig:1}
\end{figure*}

At first glance, the magnetization process of antiferromagnets is simple. In the paramagnetic regime, Langevin or Brillouin theory describes the response of independent spins $S$---treated either as classical dipoles or quantum moments---to temperature $T$ and applied magnetic field $H$~\cite{Blundell2014}. In the magnetically ordered state, assuming that spins initially lie in the plane perpendicular to the magnetic field direction $z$, the classical dipoles cannot move toward the field direction. This induces a uniform magnetization $M(H)$ which increases linearly until $\langle S^z \rangle = S$ for each spin (Fig.~\ref{fig:1}a). The onset of this fully polarized state defines the saturation field $H_{\rm sat}$. A closer inspection reveals that non-linear contributions to the isothermal magnetization encode rich phenomena. For example, quantum corrections result in a convex magnetization in Heisenberg $S\!=\!1/2$ antiferromagnets~\cite{Bonner1964, Zhitomirsky1998}. In geometrically frustrated systems, fractional magnetization plateaus can occur~\cite{Chubukov1991, Zhitomirsky2000, Matsuda2013} with $\langle \sum_i S_i^z  \rangle = \langle S^z_{\rm tot} \rangle\!=\!m\!\times\!NS$, where $m$ is a rational number and $N$ is the total number of spins (Fig.~\ref{fig:1}b). In systems with biquadratic exchange, quantum and classical effects compete in the stabilization of magnetization plateaus~\cite{Hida2005} while in frustrated $S\!>\!1/2$ systems with uniaxial anisotropy, a Devil's staircase with an infinite number of magnetization plateaus is possible at the classical level~\cite{Bak1982}. Clearly, magnetization plateaus are a widespread and fascinating occurrence in magnetic systems~\cite{Takigawa2011}. They reflect that the total spin operator is conserved and commutes with the Hamiltonian, \textit{i.e.}, $[S^z_{\rm tot},\mathcal{H}]\!=\!0$. As the resulting magnetic excitations are gapped, a reliable determination of the system Hamiltonian $\mathcal{H}$~\cite{Kamiya2018,Xie2023} is possible at much lower magnetic fields than those necessary to reach a fully polarized state. Furthermore, regardless of the underlying classical or quantum mechanisms that stabilize the plateau, the resulting magnetic state is \textit{incompressible}~\cite{Takigawa2011}; an increase in $H$ does not lead to an increase in $M(H)$. This incompressibility creates a highly sensitive platform for detecting even small amounts of chemical disorder, because spins near a defect can respond strongly to the magnetic field, \textit{i.e.}, these spins are compressible. 

To demonstrate this spectacular effect, we focus on the Ising-Heisenberg (or XXZ) triangular-lattice antiferromagnet \kcso\ (KCSO). This compound has long been known~\cite{Wildner1992,Wildner1994} and was recently revisited in detail~\cite{Zhong2020}. It features a robust magnetization plateau that spans an extraordinarily wide field range, from $2$ to $17$~T~\cite{Zhu2024}. The magnetization plateau is flanked by \textit{compressible} spin supersolid phases at low temperature~\cite{Zhu2024,Chen2024,Xu2025}. These phases are the quantum analogs of non-collinear Y and V phases, where a static magnetic order in the $z$-direction coexists with phase coherence---but no static order---transverse to $z$. Although the exact nature of these phases is not yet fully understood, we will refer to them as Y and V in the rest of this article. The crystal structure of KCSO, depicted in Fig.~\ref{fig:1}c, comprises isolated $\rm CoO_6$ octahedra connected by $\rm SeO_3$ tetrahedra, forming a triangular lattice of $\rm Co^{2+}$ ions in the $ab$-plane. These triangular layers are stacked along the $c$-axis and separated by layers of K ions. Although the effective magnetic moments of KCSO extracted from the Curie-Weiss fit at high temperatures are nearly isotropic, with $\mu_{\rm eff}^c = 6.14 \mu_{\rm B} \approx 1.13 \mu_{\rm eff}^{ab}$ (where $\mu_{\rm B}$ is the Bohr magneton) corresponding to an effective $g$-factor of $g_c=7.1$~\cite{Zhong2020}, bilinear interactions exhibit a strong Ising character. The out-of-plane nearest-neighbor exchange $J_{zz}$ is an order of magnitude larger than the in-plane exchange $J_{xy}$~\cite{Zhu2024,Chen2024}. Above the critical field $\mu_0H_{\rm c1}\approx2$~T (where $\mu_0$ is the vacuum permeability) and at or below $T=5$~K, KCSO stabilizes an archetypal up-up-down (UUD) phase associated with a magnetization plateau at $m\!=\!1/3$. This state persists until the critical field $\mu_0H_{\rm c2}\approx17$~T.

\begin{figure*}[htb!]
        \centering
        \includegraphics[width=1.0\linewidth]{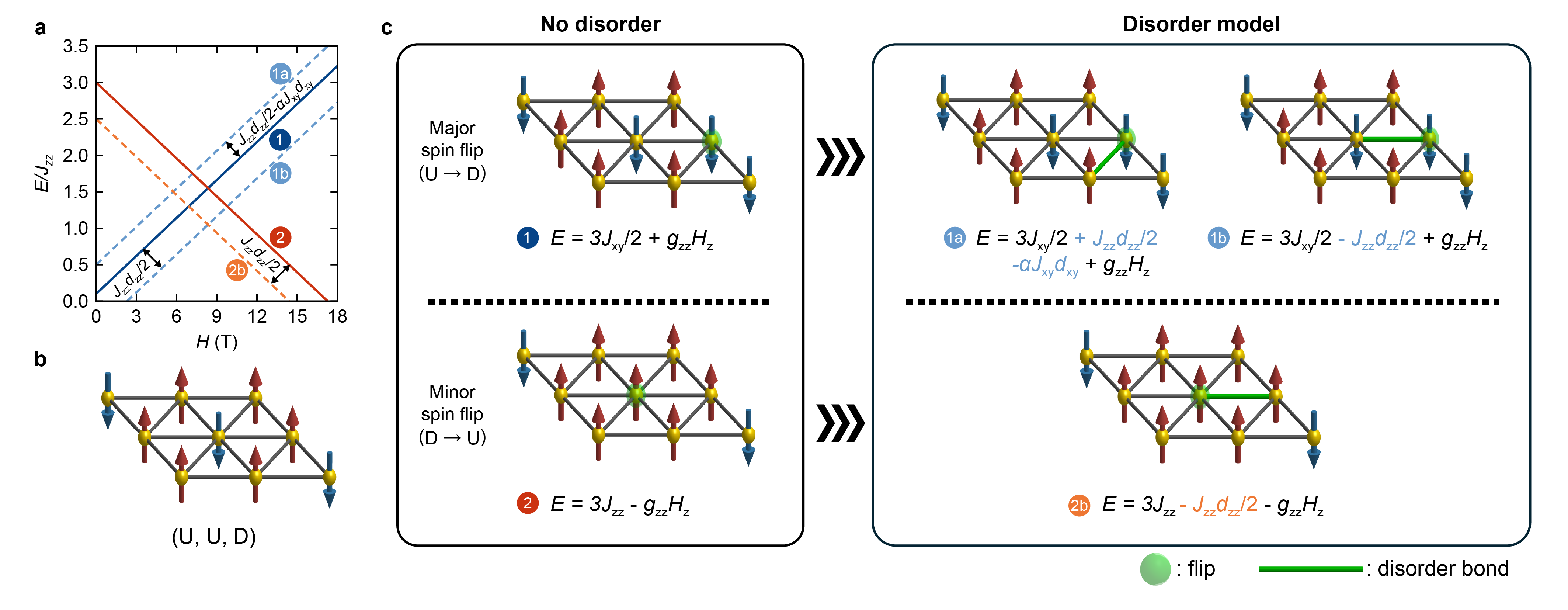}
        \caption{\textbf{Schematic interpretation of magnetic excitations in the UUD phase of \kcso.} \textbf{a, } Magnetic excitations expected in the UUD phase with and without bond disorder. Solid lines (labeled \#1 and \#2) represent excitations in a clean system, while dashed lines (labeled \#1a for above, and \#1b, \#2b for below) correspond to additional excitations in a disordered system. Blue and light blue indicate major spin-flip processes, while red and orange represent minor spin-flip processes. \textbf{b, } Spin structure of the ideal UUD phase for a $3\times3$ triangular-lattice supercell. \textbf{c, } Visualization of major and minor spin-flip processes in clean and disordered systems. The flipped spin in each process is shown with a green sphere. The green bond represents a disorder bond with modified exchange interactions $J_{zz}’ = J_{zz}(1-d_{zz})$ and $J_{ xy}’ = J_{xy}(1-d_{xy})$.}
        \label{fig:2}
\end{figure*}

\section*{Results}

Here, we utilize far-infrared magneto-optical spectroscopy (FIRMS) and thermomagnetic measurements to investigate the behavior of KCSO near and within the plateau at $m=1/3$ (see Methods). To our surprise, although our crystals are grown using published protocols and chemical disorder ruled out by advanced X-ray diffraction and electron microscopy methods (see Supplementary Sec. V and VI
), the measurements reveal sharp spectroscopic fingerprints that can be clearly linked to very low levels of chemical disorder. These fingerprints were not seen in previous measurements~\cite{Chen2024}. Our optical measurements in the UUD phase reveal two dominant modes accompanied by three satellite modes, as illustrated in Fig.~\ref{fig:1}d and \ref{fig:1}e. These satellite modes cannot be explained by clean-system models across three simulation techniques: linear spin wave theory (LSWT), Landau-Lifshitz dynamics (LLD), and exact diagonalization (ED). Instead, we develop a disordered-system model that fully captures our experimental observations. In this framework, the energy shift between the dominant and satellite modes reflects the nature and strength of exchange disorder on nearest-neighbor bonds. The large and discrete energy shifts in our data suggest that the disorder in our crystals is akin to missing bonds or spin vacancies, rather than a distribution of weakened bonds. As we illustrate below, a typically small amount of vacancies---as little as 2.3\%---is sufficient to produce the observed intensity of the satellite modes. Moreover, magnetization and heat capacity measurements reveal a secondary magnetization plateau state with $m\!<\!1/3$, which is successfully reproduced in classical Monte Carlo (MC) simulations using the vacancy model.

Our FIRMS results, measured in Faraday geometry ($H \parallel c \equiv z$) at $T\!=\!5$~K, are shown in Fig.~\ref{fig:1}d. At low magnetic fields ($0<\mu_0H<3$~T), the system is in the Y phase~\cite{Chen2024}, and the spectra exhibit two broad field-dependent modes that cross at $\mu_0 H \approx$ 3 T. Around 3 T, the system enters the UUD phase and displays much sharper modes, with the two strongest intersecting at $\mu_0H\!=\!9.75$~T. Three satellite modes, running parallel to the main modes, are also observed. Above 15 T, an additional mode emerges and broadens with increasing field; we associate this mode with the onset of the V phase~\cite{Chen2024}. We determine the energy of major modes at each magnetic field, $E(H)$, using Lorentzian fits, and plot the results on Fig.~\ref{fig:1}e, where the color encodes the parent Y, UUD, or V phase associated with each mode. Now, we address the origin of the two strongest modes in the UUD phase. The magnetic Hamiltonian of KCSO in that field configuration reads
\begin{eqnarray}
    \mathcal{H} = && \sum_{\langle i,j \rangle}[J_{xy}(S_{i}^xS_{j}^x+S_{i}^yS_{j}^y)+J_{zz}S_{i}^zS_{j}^z] \nonumber \\ && + \mu_0\mu_{\rm B} g_{z}H\sum_iS_{i}^z,
    \label{eq:Hamiltonian}
\end{eqnarray}
where the first sum runs over nearest neighbor bonds, $J_{xy}>0$ and $J_{zz}>0$ are antiferromagnetic interactions, and $S_{i,j}$ are spin-1/2 operators. The second term describes the usual Zeeman effect, with magnetic field $H$ in the $z$-direction, and $g_{z}$ representing the $z$-component of the effective $g$-factor for the Co$^{2+}$ ions. In the classical Ising limit, the excited states of the UUD phase can be expressed as integer multiples of $J_{zz}$:
\begin{eqnarray}
    E_{n}=nJ_{zz} \pm \mu_0\mu_{\rm B} g_{z} H,~~(n=0,1,2,3,…).  
    \label{eq:disorded_bond}
\end{eqnarray}

Due to optical selection rules, only transitions with $\Delta S^z = \pm1$ are visible in FIRMS. Therefore, our experiment primarily detects excited states involving single or triple spin-flip processes (see Supplementary Sec. II 
for further discussion). The two strongest modes observed in the UUD phase can be naturally attributed to single spin-flip excitations. In Fig.~\ref{fig:2}a, solid lines represent the mode energies $E(H)$ expected for single spin-flips allowed by the selection rule. These processes can occur at two distinct spin sites: up spin (major spin) and down spin (minor spin). A minor spin flip costs onsite energy in terms of $J_{zz}$ only---specifically, $3J_{zz}$ for the triangular lattice---since the flipped spin cannot propagate to neighboring spin sites. In contrast, a major spin flip incurs no onsite energy change in the classical limit but is accompanied by an additional quantum correction of $3J_{xy}/2$, reflecting that the flipped spin can propagate to three neighboring up-spin sites (Fig.~\ref{fig:2}b,c). Linear fits to these two modes (see Supplementary Table 2
) yield $J_{xy}$ = 0.20(1) meV, $J_{zz}$ = 2.88(1) meV, and $g_{z}$ = 7.45(5), in good agreement with previous neutron scattering and magnetization studies~\cite{Zhong2020,Zhu2024,Chen2024}, but providing a more accurate determination of the $g$-factor. Remarkably, in the UUD phase, all main and satellite modes appear resolution-limited, as shown in Supplementary Figure 4. Triple spin-flip excitations, forming both a bound-state and a free particle continuum, are also observed (see Supplementary Sec. IIA for details).

Our data in the UUD phase exhibit three salient satellite modes that do not correspond to either single or triple spin-flip processes. To account for these excitations, we develop a disorder-bond model in which exchange interactions are modified as:
\begin{eqnarray} \label{eq:disorded_bond}
    J_{zz}' =J_{zz}(1-d_{zz}) , \quad
    J_{xy}' = J_{xy}(1-d_{xy}),   
\end{eqnarray}
where $d_{zz}$ and $d_{xy}$ represent the disorder-induced reduction of the respective interaction. Notably, this model yields fine spectroscopic features that are the unmistakable hallmark of the disorder's nature. Specifically, a major spin-flip excitation can split into two modes, depending on whether the disorder bond is connected to a major spin (mode \#1a in Fig.~\ref{fig:2}) or a minor spin (mode \#1b in Fig.~\ref{fig:2}). The excitation energies are shifted by $\pm J_{zz}d_{zz}/2$, respectively. Moreover, when the disorder bond is connected to a major spin, there is a quantum correction of $-\alpha J_{xy}d_{xy}$ (where $\alpha \simeq n_{dm}$ is the disorder density in the dilute limit), due to the propagation of the flipped spin to two neighboring up-spin sites with intact bonds. In the case of a minor spin-flip process, the disorder only introduces an energy shift of $-J_{zz}d_{zz}/2$, since the flipped spin cannot propagate to neighboring spin sites (see detailed discussion in Supplementary Sec. II
D). More generally, we hypothesize that the splitting and multiplicity of spin-flip excitations in incompressible magnetic states serve as a unique indicator of bond disorder.

\begin{figure}[htb!]
        \centering
        \includegraphics[width=0.95\linewidth]{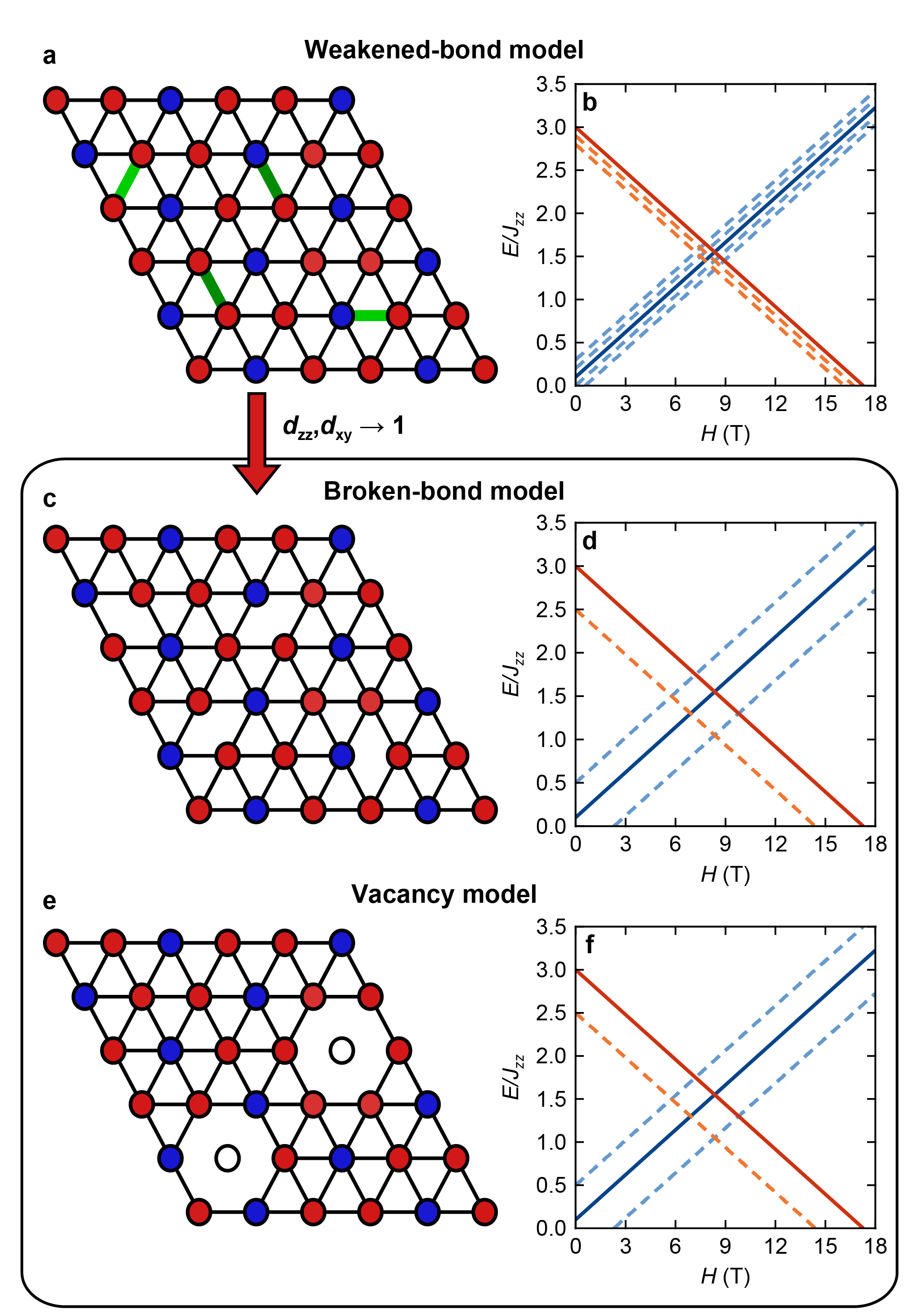}
        \caption{\textbf{Universality of the disorder-bond model and its correspondence with the vacancy model.} Schematic UUD spin configurations and excitation modes for: \textbf{a-b, } the weakened-bond model, where for illustration the exchange interactions on green and dark green bonds are reduced with parameters $d_{zz}=d_{xy} =0.1 $ (green) and $d_{zz}=d_{xy} = 0.2$ (dark green); \textbf{c-d, } the broken-bond model, where $d_{zz}=d_{xy} \rightarrow 1$ for the same broken bond; and \textbf{e-f, } the vacancy model, where two magnetic sites are removed. Throughout, red (blue) circles indicate up (down) spins. The line style for the spectra in \textbf{b}, \textbf{d}, and \textbf{f} follows the convention of Fig.~\ref{fig:2}. The spectral intensity of the satellite modes is not shown and differs between panels \textbf{d} and \textbf{f}, which represent different defect densities.}
        \label{fig:3}
\end{figure}

\begin{figure*}[htb!]
        \centering
        \includegraphics[width=1.0\linewidth]{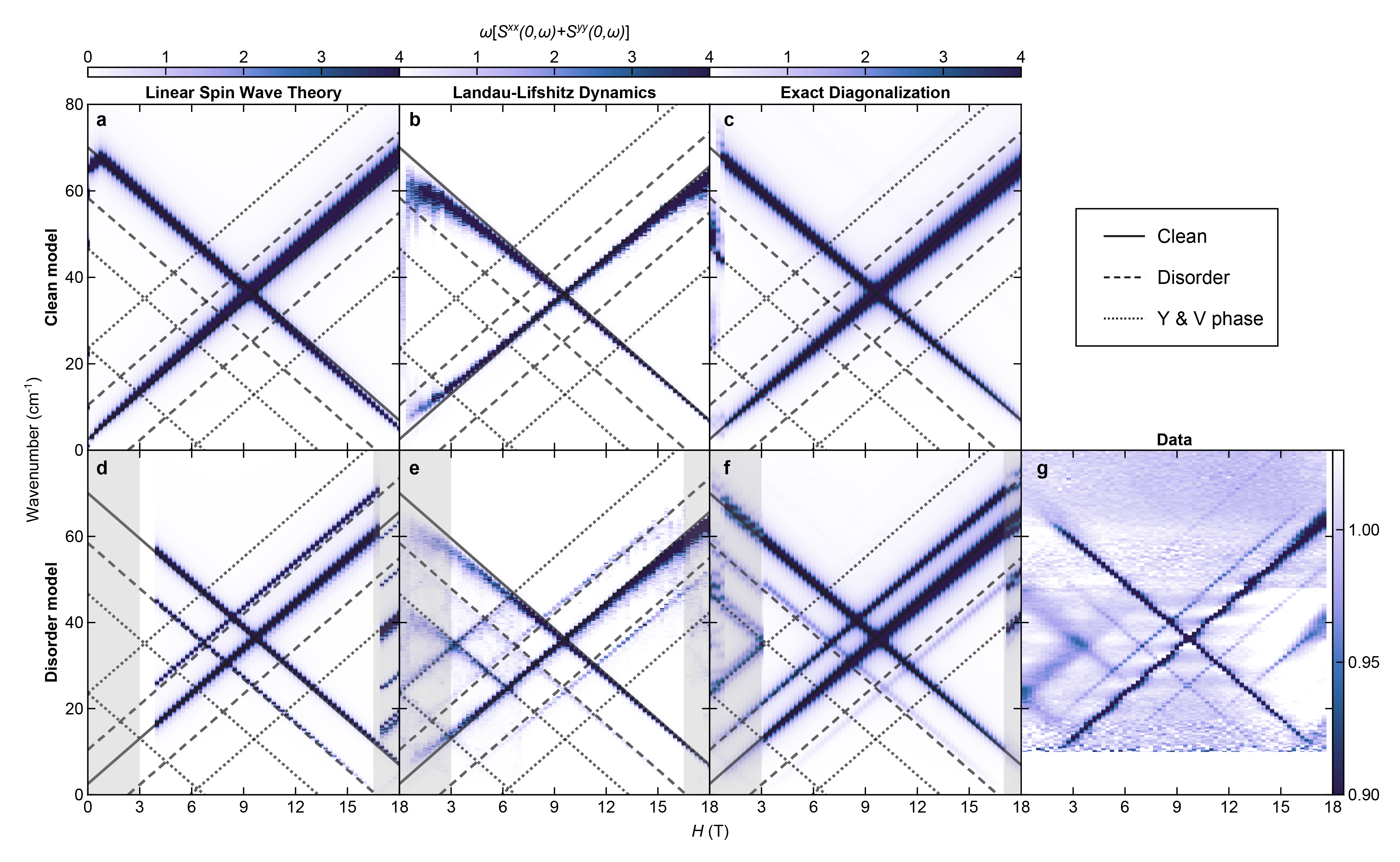}
        \caption{\textbf{Simulated FIRMS spectra from our model of \kcso\ with and without disorder.} \textbf{a-c, } Frequency and magnetic field dependence of simulated $I(\omega; H)$ for a clean system simulated using LSWT, LLD, and ED, respectively. \textbf{d-f, } Similar simulation approaches but for a disordered system. Gray lines represent the linear fits to mode energies, identical to those reported in Fig.~\ref{fig:1}e. Gray areas indicate the regime where the ground state of the system is not UUD. \textbf{g, } Same data as in Fig.~\ref{fig:1}d, shown for better comparison with the simulation results.}
        \label{fig:4}
\end{figure*}

\begin{figure*}[htb!]
        \centering
        \includegraphics[width=1.0\linewidth]{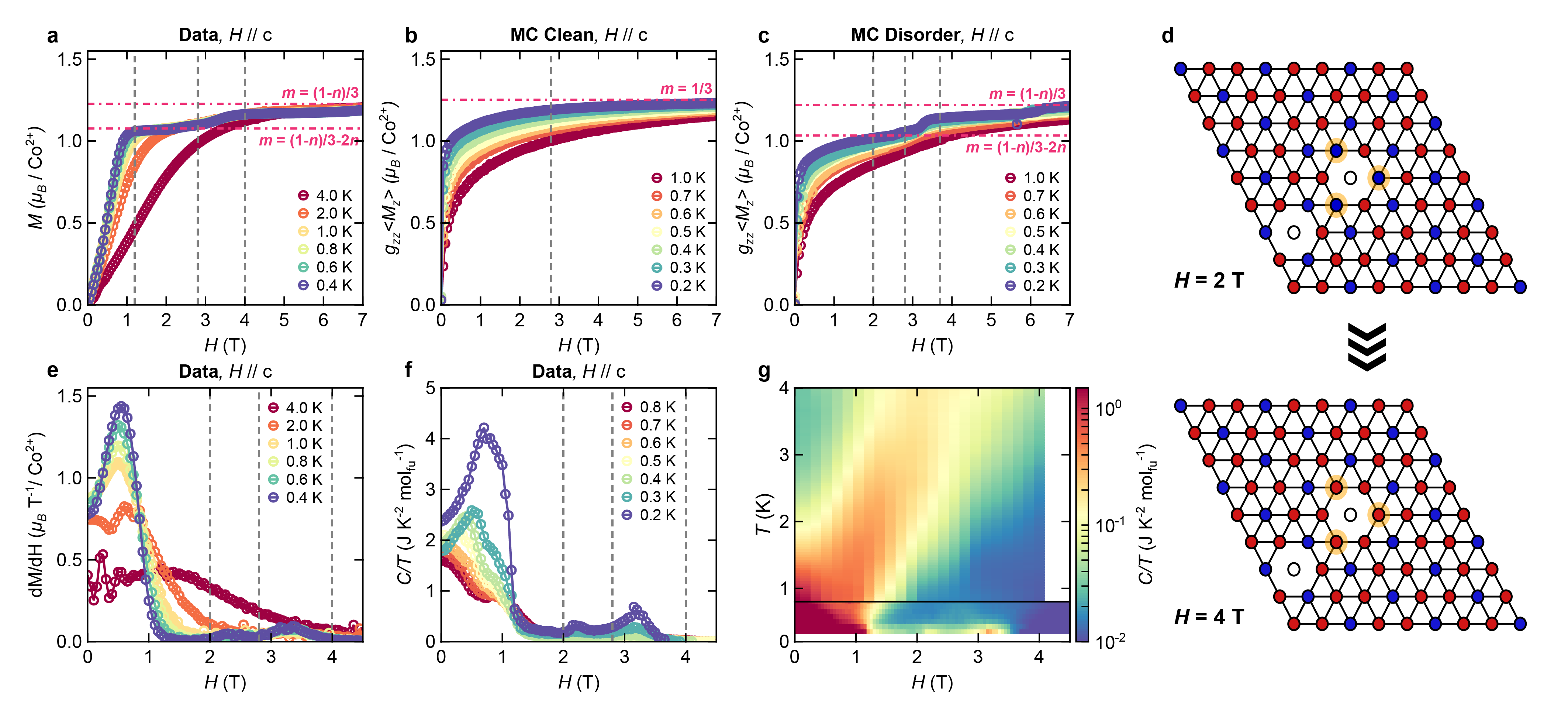}
        \caption{\textbf{Magnetization and specific heat of our crystals of \kcso.} \textbf{a, } Temperature dependence of the magnetization of KCSO in a $c$-axis magnetic field. \textbf{b-c, } Simulated isothermal magnetization curves using the classical MC method for clean and disordered systems, respectively. Magenta dot-dashed lines in each figure indicate the magnetization of the pre-plateau and UUD plateau phase with the vacancy concentration $n$. \textbf{d, } Schematic mechanism stabilizing an additional magnetization plateau around $\mu_0H =2.8$~T. The top and bottom panels show the spin configurations at 2 T and 4 T, respectively. Red, blue, and white circles indicate up spins, down spins, and vacancies, respectively. Orange circles highlight spin-flip sites that result in the additional plateau. \textbf{e,} Field derivative of the magnetization. \textbf{f-g, } Temperature and magnetic field dependence of the specific heat divided by temperature for KCSO in a $c$-axis magnetic field. The colorplot combines constant-field ($T\geq 0.8$~K) and constant-temperature ($T < 0.8$~K) measurements.}
        \label{fig:5}
\end{figure*}

Using the analytical solution of the disorder-bond model, we perform linear fits to the observed satellite modes (see Supplementary Sec. I
C). We find that the $y$-intercepts of these three modes are consistent with the model prediction, as illustrated in Fig.~\ref{fig:2}a. Quantitatively, we obtain $d_{zz}\!=\!0.93$ from the energy shift between modes \#1 and \#1b, and $d_{zz}=0.96$ from the energy shift between modes \#2 and \#2b. The close-to-unity value for $d_{zz}$ points to the complete absence of a bond, rather than a mere modification of the exchange due to disorder. The simplest mechanism to generate missing bonds is through spin-site vacancies, \textit{i.e.}, missing $\rm Co^{2+}$ ions.

The disorder-bond framework of Eq.~\ref{eq:disorded_bond} interpolates between three distinct models of random disorder: a weakened-bond model, a broken-bond model, and a vacancy model. Figure~\ref{fig:3} schematically illustrates the spin and exchange configurations for each case and the associated excitation spectra. In the weakened-bond model, multiple bonds exhibit randomly distributed reduction factors $d_{zz}$ and $d_{xy}$. The resulting satellite excitations detach from the main modes, with their energy offsets set by the reduction factors. Consequently, for a distribution of weakened bonds, we expect the main spectral lines to be broadened and sharp satellite modes to be absent, depending on the details of the disorder distribution (see Fig.~\ref{fig:3}b). By contrast, in the broken-bond model, the reduction factors are fixed to unity on each defective bond. This produces sharp satellite modes well separated from the main modes, which also remain sharp (see Fig.~\ref{fig:3}d). Remarkably, the vacancy model yields an identical spectrum, albeit with distinct intensity, because the six bonds adjacent to a vacancy are effectively removed (see Fig.~\ref{fig:3}e). Thus, despite their distinct chemical origins, the broken-bond and vacancy models are \textit{spectroscopically indistinguishable} for an unknown defect density. From a materials standpoint, however, the complete absence of a specific bond is less realistic than the presence of a magnetic-ion vacancy.

Based on the above analysis and exchange parameters, and the fact that all main and satellite modes in the UUD phase are resolution limited, we simulated the FIRMS spectra with and without vacancy disorder. We employed three different approaches, as summarized in Fig.~\ref{fig:4}: LSWT, LLD (to account for finite temperature), and ED (to account for quantum corrections). Calculations for a clean system, Fig.~\ref{fig:4}a-c, show that two sharp single spin-flip modes entirely dominate in the $m\!=\!1/3$ UUD phase, with only weak continua of triple spin-flip excitations in ED at high energies (see Supplementary Sec. II
A). Simulations including disorder paint a different picture, Fig.~\ref{fig:4}d-f. To account for disorder in LSWT and LLD, we employ a spin vacancy model with non-adjacent vacancies in supercells up to size $18 \times 18$. In ED, due to system size constraints, we harness the spectroscopic equivalence between broken bond and vacancy models and use the former with $d_{zz}=1$ for two bonds in a $6 \times 3$ supercell, which equivalently reproduces the same excitations as the vacancy model (see Supplementary Sec. II
C). As the vacancy/disorder density is directly related to the intensity ratio between the dominant and satellite excitations, we vary the number of defects (in multiples of 3) in the ($18\times18$) supercell until the simulations are in excellent agreement with the data (Supplementary Sec. IIB). We find that 2.3(8)\% of $\rm Co^{2+}$ vacancies is the most likely range of concentrations, as it successfully reproduces the intensity of all satellite modes in the UUD phase. A more naive analytical approach to estimate the vacancy concentration (Supplementary Sec. IIE), which neglects spectroscopic matrix elements, yields 3.2(14)\% of vacancies, within error bars of the more elaborate spin-wave-based method.

We now examine the impact of disorder on the thermomagnetic properties of our crystals of KCSO. Typical thermodynamic measurements are sensitive enough to quantify the impurity concentrations in diluted spin systems \cite{Nikitin2020, Cairns2022}, and here we extend this concept to a dense spin system. The isothermal magnetization $M(H)$ measured to sub-Kelvin temperatures with the field along the $c$-axis is shown in Fig.~\ref{fig:5}a. In the clean case, the magnetization is expected to reach the $m=1/3$ plateau in a single step (see Fig.~\ref{fig:1}b for a cartoon), as observed by previous experimental studies~\cite{Zhu2024,Chen2024} (see also Supplementary Sec. III
). In our case, however, an additional plateau with $m<1/3$ is observed from $\mu_0H\approx2.8$~T, and the main plateau never reaches exactly $m=1/3$. To understand this behavior, we performed classical MC simulations for both our clean and disordered models of KCSO with a 2.5\% vacancy concentration close to the representative value obtained from FIRMS measurements. In the clean case, the magnetization saturates at $m=1/3$ near $\mu_0H\approx4$~T (Fig.~\ref{fig:5}b). In contrast, the magnetization in the disordered case shows a pre-plateau around $\mu_0H\approx2.8$~T, followed by a jump to the primary plateau at 3.4 T (Fig.~\ref{fig:5}c). The magnetization on both plateaus falls below $m=1/3$. By tracking the spin configuration at each magnetic field in our MC simulations, we find that the pre-plateau originates from spin flips near vacancy sites (Fig.~\ref{fig:5}d). With this understanding, our naive analytical approach (Supplementary Sec. IIE) can be used to estimate the value of the normalized magnetization per spin. In presence of a vacancy concentration $n$, it yields $(1-n)/3$ and $(1-n)/3-2n$ for the magnetization at the main plateau and pre-plateau, respectively. This approach best matches our magnetization measurements for $n=2.0\%$ (Fig.~\ref{fig:5}a), which is remarkably close to the best concentration inferred from FIRMS results, giving further confidence in our vacancy-based disorder model. Thus, the additional plateau we observed serves as a thermodynamic signature of disorder. Taken independently, it lacks the mechanistic insights provided by our spectroscopic approach. But with insight about the disorder mechanism, it yields a surprisingly accurate determination of the disorder concentration.

Heat capacity measurements at low temperatures, with the magnetic field applied along the $c$-axis, are presented in Fig.~\ref{fig:5}f-g. The field dependence of specific heat at $T=0.2$~K reveals that a phase with gapless excitations exists below $\mu_0H\approx1$~T, which we assign to the Y-phase or its quantum analog. As the field increases, the specific heat is suppressed but reaches a finite value again between $\mu_0H\approx2.1$~T and 3.5~T (Fig.~\ref{fig:5}f). This enhancement survives up to $T\!=\!0.5$~K and matches the peaks observed in the field derivative of magnetization (Fig.~\ref{fig:5}e). This correlation suggests that the low-temperature specific heat can detect the disorder-induced plateau state through the density of states of local spin-flip processes near vacancies, as these become gapless. The value of the entropy release between $T\!\approx\!0.2$ and $0.8$~K near $\mu_0H\approx3$~T is consistent with local spin-flip excitations near vacancies (see Supplementary Sec.~IV). Above $T\!=\!1$~K, the heat capacity shows only a broad peak that corresponds to the field-driven transition from the Y phase to the UUD phase. These findings emphasize the significance of examining the low-temperature regime when investigating the effects of disorder, even in systems that exhibit Ising-limit behavior.

\section*{Discussion}

The spectacular spectroscopic signatures of disorder we observe in the $m\!=\!1/3$ plateau phase are not unique to KCSO. Additional measurements of the sister compound $\rm Rb_2Co (SeO_3)_2$ (see Supplementary Sec. I
E) reveal similar satellite modes. Notably, the 2.3(8)\% vacancy concentration extracted in this work appears below the detection limit of laboratory diffraction and microscopy techniques commonly used in the routine structural characterization of quantum materials. In the UUD phase, this density of vacancies does not jeopardize the original magnetic ground state, unlike other systems where the true ground state of the system may be concealed by disorder effects \cite{pustogow2020, itou2023, freedman2010site}. Further studies are needed to elucidate if and how the gapless phases of KCSO, such as the V- and Y- phases (and their supersolid counterparts) are affected by the disorder we uncovered.

By combining the physics of an incompressible magnetization plateau state with the high sensitivity of magneto-optical measurements, we demonstrate a viable route to probe disorder in magnetic insulators. This approach not only provides quantitative insights into the amount of disorder, but also informs about its mechanistic origin. Given that magnetization plateaus are a common occurrence in frustrated and quantum magnets, and that magneto-optical techniques are becoming more accessible in laboratories worldwide, our method has the potential to accelerate the search for and validation of emergent magnetic states, such as QSLs.

\section*{Methods}

\subsection*{Sample preparation}
Single crystals of \kcso\ were prepared following the procedure reported in Ref.~[\onlinecite{Zhong2020}]. Specifically, high-purity and commercially available K$_2$Co$_3$, Co$_3$O$_4$, and SeO$_2$ were mixed in a molar ratio of $1.2\!:\!0.33\!:\!2.2$ and loaded into an Al$_2$O$_3$ crucible, which was sealed in a vacuumed quartz tube. The tube was heated to $T\!=\!600~^{\circ}$C, maintained at that temperature for 8 hours, and then cooled to room temperature at a rate of $6~^{\circ}$C/hour. The resulting flat pink crystals were collected in water. The crystallographic $c^\ast$ and $c$ axes are normal to the flat surface of the crystals, as confirmed by backscattering Laue X-ray diffraction. The largest K$_2$Co(SeO$_3$)$_2$ crystal was selected for far-infrared magneto-spectroscopy (FIRMS) measurements and
was polished to create parallel planes perpendicular to the $c$-axis. Two unpolished crystals were used for thermodynamic measurements, which yielded consistent results. Single crystals of Rb$_2$Co(SeO$_3$)$_2$ were prepared similarly.

\subsection*{Far-infrared magneto-spectroscopy}
FIRMS measurements were performed using a Bruker VERTEX 80v Fourier-transform infrared spectrometer at the National High Magnetic Field Laboratory (NHMFL, Tallahassee, Florida). The sample was mounted with GE varnish on a nonmagnetic aperture (1.5~mm in diameter) and attached to a top-loading probe. The probe was then inserted into an optical cryostat equipped with a superconducting magnet reaching up to $\mu_0 H\!=\!17.5$~T. Infrared light from a mercury lamp was delivered to the sample via an evacuated light pipe and focused onto the sample surface using a parabolic cone.

To reduce light-induced heating and improve signal-to-noise ratio, a 700~cm$^{-1}$ low-pass filter was placed in the light path just above the sample. The transmitted infrared light passing through the sample was collected by another pair of parabolic cones and detected by a Si bolometer operating at liquid helium temperature. A schematic of the resulting Faraday transmission configuration, with the light propagating along $H\!\parallel\!c$, is shown in Supplementary Fig. 1
. The lowest effective sample temperature during measurements was $T_{\rm min} = 5$~K, as measured by a thermometer mounted on the sample holder.

During the measurements, three different beam splitters (Multilayer Mylar, Mylar 50~$\mu$m, and Mylar 125~$\mu$m) were used to cover the relevant energy range. The raw spectra from each (beam splitter) measurement are shown in Supplementary Fig. 2
.

\subsection*{Spin dynamics simulation}
Simulations of the transmitted frequency-dependent FIRMS spectra $I(\omega)$ were performed using the {\scshape Sunny.jl} package~\cite{Sunny2025} and the {\scshape QuSpin.py} package~\cite{QuSpin-1,QuSpin-2} by calculating the diagonal components of the frequency-weighted zero-momentum-transfer dynamical spin structure factor perpendicular to the light transmission direction,
\begin{eqnarray}
    I(\omega) = \omega[S^{xx}({\bf q} = 0, \omega) + S^{yy}({\bf q} = 0, \omega)],
    \label{eqsi:clss}
\end{eqnarray}
where $z\!\parallel\!H \!\parallel\!c$.
The dynamical spin structure factor was simulated using several approaches; in either case, magnetic moments were treated as dipolar spins and the system's dynamics was governed by the same parent quantum Hamiltonian $\mathcal{H}$, Eq. \ref{eq:Hamiltonian}, and its known exchange parameters. 

The zero-temperature response ($T\!=\!0$~K) was simulated using linear spin wave theory, by which spin-1/2 operators ${\bf S}_i$ are transformed into Holstein-Primakoff bosons ${b}_i$ and the resulting bosonic Hamiltonian, truncated at quadratic order, is diagonalized using known numerical approaches implemented in {\scshape Sunny.jl} and other programs~\cite{Toth2015,Petit2016}.

The finite-temperature response was obtained in {\scshape Sunny.jl} using classical Landau-Lifshitz dynamics (LLD). For these simulations the spin system is assumed to be a product state $|{\bf \Omega} \rangle = \otimes_i |{\bf \Omega}_i\rangle$ over SU(2) coherent states $|{\bf \Omega}_i\rangle$ representing dipolar spin operators $\hat{{\bf S}}_i$, which are replaced with classical vectors ${\bf \Omega}_i = \langle  {\bf \Omega}_i | \hat{\bf S}_i |{\bf \Omega}_i \rangle$. The time-dependent dynamics at finite temperature is calculated using the stochastic Landau-Lifshitz-Gilbert (LLG) equation:
\begin{eqnarray}
    \frac{d\mathbf{\Omega}}{dt} = -\mathbf{\Omega}\times\left[\mathbf{\xi}+\frac{d \mathcal{H}_{\rm cl}}{d\mathbf{\Omega}}-\lambda\left(\mathbf{\Omega}\times\frac{d \mathcal{H}_{\rm cl}}{d\mathbf{S}}\right)\right],
    \label{eq:stochastic_LLG_eq}
\end{eqnarray}
where $\mathcal{H}_{\rm cl} =  \langle {\bf \Omega} | \mathcal{H} | {\bf \Omega} \rangle$ in the large-$S$ limit, $\mathbf{\xi}$ is a temperature-dependent Gaussian white noise, and $\lambda = 0.2$ sets the coupling strength between the system and the thermal bath. The simulations were performed using a $9\times9$ super-cell of the conventional chemical unit cell of the crystal with periodic boundary conditions.  To match the overall excitation bandwidth of our data, we set the temperature to $T = 0.5$~K in the LLD simulations. The system was first thermalized over 5,000 Langevin time steps with a step size of $\Delta t = 0.013$ meV$^{-1}$. After thermalization, five spin configurations were sampled using the LLG equation, with each configuration separated by 3,000 Langevin time steps to ensure decorrelation.

Exact diagonalization calculations were performed using {\scshape QuSpin.py} \cite{QuSpin-1, QuSpin-2}. The frequency-weighted dynamical spin structure factor, Eq.~\ref{eqsi:clss}, was calculated using a $6\times3$ supercell with periodic boundary conditions, employing the Lanczos algorithm \cite{lanczos1950iteration} and the continued fraction method \cite{Continuedfraction1994}. The number of Lanczos iterations was set to 200. Due to computational limitations, quenched chemical disorder was modeled using weak bonds rather than directly introducing vacancies. A more detailed discussion can be found in the Supplementary Sec. II
C.

\subsection*{Classical Monte Carlo simulation}
Field-dependent magnetization was calculated from the standard Monte Carlo Metropolis algorithm using {\scshape Sunny.jl} package \cite{Sunny2025}. The supercell size for the Monte Carlo simulation was set to $18 \times 18 \times 1$. For each magnetic field, we perform 20000 spin-flip steps for the annealing at a given temperature. After the annealing process, 20 samples of spin configurations are averaged after 100 spin flips for each sample. 

\subsection*{Thermomagnetic measurements}
Isothermal magnetization measurements were performed using a Quantum Design MPMS-3 SQUID magnetometer operated in DC mode with the $^3$He insert option (temperatures from $T=0.4$~K to 4~K). The magnetic field was applied along the $c$-axis. Heat capacity measurements were conducted using a Quantum Design Dynacool PPMS equipped the Dilution Refrigerator unit. Magnetic fields up to $\mu_0 H =5$~T were also applied along the $c$-axis to map the temperature- and field-dependent specific heat $C(T,H)$ using the thermal relaxation method over a temperature range from $T=200$~mK to 4~K. Measurements above $T=800$~mK were performed at fixed magnetic fields with a high density of data points along the temperature axis, while measurements below  $T=800$~mK were conducted at fixed temperatures with a higher density of data points along the magnetic field axis.

\subsection*{Single-crystal X-ray diffraction}
Single-crystal X-ray diffraction (SCXRD) was performed on a specimen with dimensions $0.15\times0.1\times0.08$~mm$^3$. The crystal was affixed to a nylon loop using Paratone oil and examined with a Rigaku XtaLAB Synergy, Dualflex, Hypix SCXRD diffractometer, operated at room temperature.

Crystallographic data were acquired using the $\omega$-scan method with Mo K$\alpha$ radiation ($\lambda = 0.71073$~Å) from a micro-focus sealed X-ray tube operating at $\Delta V=50$~kV and $I=1$~mA. Experimental parameters, including the total number of runs and images, were determined algorithmically based on strategy computations performed by CrysAlisPro software (version 1.171.42.101a, Rigaku OD, 2023). A detailed analysis of the SCXRD data is provided in Supplementary Sec. V
.

\subsection*{Scanning transmission electron microscopy}
A focused ion beam from a Thermo Fisher Helios 5CX was used to lift out a specimen from a \kcso\ single crystal. The specimen was then polished to a thickness of $<50$~nm to achieve atomic resolution using a Hitachi HD-2700 scanning transmission electron microscope (STEM). The microscope was operated at $\Delta V=200$~keV with a convergence angle of 27~mrad and a spatial resolution of $\Delta r\approx 1.3$~\AA. Annular dark-field (ADF) images were collected and compared with the \kcso\ lattice structure to determine the zone axis. A selected ADF image is shown in Supplementary Fig. 10
.

\begin{acknowledgements}
We thank N. Peter Armitage for insightful discussions. This work (spectroscopy, thermomagnetic measurements, data analysis, modeling, interpretation and theory) was primarily supported by the U.S. Department of Energy, Office of Science, Office of Basic Energy Sciences under grants DE-FG02-07ER46451 (SR, NZ, NVS, DS, and ZJ), DE-SC0018660 (CK and MM), and Early Career Research Program DE-SC0025478 (theory work by AS and IK). The crystal growth effort at UTK (AR, LC, and HZ) was supported by the U.S. Department of Energy grant DE-SC0020254. The SCXRD measurements at MSU (CP, MX, and WX) were supported the U.S. Department of Energy grant DE-SC0023648. STEM measurements were performed at the GT Institute for Matter and Systems, a member of the National Nanotechnology Coordinated Infrastructure (NNCI), which is supported by the National Science Foundation under grant NSF-ECCS-2025462. FIRMS measurements were performed at the NHMFL, which is supported by the National Science Foundation Cooperative Agreement under grant NSF-DMR-2128556 and the State of Florida. 
\end{acknowledgements}

\section*{Author Contributions}
SR, NZ, NVS., MO, and DS performed the FIRMS measurements. CK, HZ, and MM carried out the thermodynamic measurements. CK, SR, MM, and ZJ analyzed the experimental data, while CK performed the spin dynamics simulations under the guidance of IK, MM, and ZJ. AS and IK proposed and analyzed the disorder-bond model. AR, LC, and HZ synthesized the single crystals. CP, MX, and WX performed the SCXRD measurements and fitting. ADV and MT conducted the STEM measurements. CK, SR, MM, and ZJ wrote the manuscript with input from all coauthors. ZJ and DS supervised the project.

\section*{Competing interests}
The authors declare no competing interests.

\section*{Data availability}
The data that support the findings of this study will be available during peer review in the Georgia Tech public data repository at: \url{https://hdl.handle.net/1853/72971}.

\section*{Codes availability}
The codes that support the findings of this study are available from the corresponding authors upon reasonable request. 

\bibliography{main_text_ref}

\end{document}